\documentclass[prd,aps,twocolumn]{revtex4}
\usepackage{latexsym}
\usepackage{amssymb}
\usepackage[dvips]{graphicx}
\begin{document}
\newcommand{\beq}{\begin{equation}}
\newcommand{\eeq}{\end{equation}}
\newcommand{\ove}{\overline}
\newcommand{\half}{\frac 1 2 }
\newcommand{\fourth}{\frac 1 4}
\newcommand{\Fstar}{\raisebox{.2ex}{$\stackrel{*}{F}$}{}}
\newcommand{\Pstar}{\raisebox{.2ex}{$\stackrel{*}{P}$}{}}
%
\newcommand{\et}{{\em et al}}
\newcommand{\ie}{{\em i.e.$\;$}}
%
%
\newcommand{\Prd}{Phys.  Rev. D$\;$}
\newcommand{\Prl}{Phys.  Rev.  Lett.}
\newcommand{\Plb}{Phys.  Lett.  B}
\newcommand{\Cqg}{Class.  Quantum Grav.}
\newcommand{\Np}{Nuc.  Phys.}
\newcommand{\Grg}{Gen.  Rel.  \& Grav.}
\newcommand{\Fp}{Fortschr.  Phys.}
\newcommand{\Sch}{Schwarszchild$\:$}
\renewcommand{\baselinestretch}{1.2}

\title{Production of photons in a bouncing universe}

\author{J. M. Salim$^1$, S. E. Perez Bergliaffa$^{1,2}$ and N. Souza$^2$}
\affiliation{$^1$Centro Brasileiro de Pesquisas Fisicas, Rua
Xavier Sigaud, 150, CEP 22290-180, Rio de Janeiro, Brazil,}
\affiliation{$^2$ Departamento de F\'{\i}sica Te\'{o}rica,
Instituto de F\'{\i}sica, Universidade do Estado de Rio de Janeiro,
CEP 20550-013, Rio de Janeiro, Brazil
}
\vspace{.5cm}

\begin{abstract}
Using a new non-singular solution, it is shown that the production
of photons in dilaton electrodynamics in a cosmological setting is
increased if the effect of matter creation on the geometry is
taken into account. This increment may be related to the problem
of the origin of magnetic fields in the universe.
\end{abstract}

\vskip2pc
\maketitle

\section*{Introduction}

The origin, evolution, and structure of the magnetic fields
present in galaxies and clusters of galaxies
are amongst the most important
open issues in astrophysics and cosmology.
Typically, the magnetic field present in galaxies is of the order of a few $\mu$G,
and is coherent
on the galactic scale. The observed field in the case of clusters is correlated over
10-100 kiloparsecs, and can be as high as tens of $\mu$G.
The standard mechanism to account for these fields
is the dynamo\cite{parker}, which amplifies small seed fields to the
abovementioned values \cite{kul}.
The seeds that fuel the dynamo may have
an astrophysical origin, or may be primordial.
An astrophysical mechanism
for the generation of these pre-galactic magnetic fields
is
the Biermann battery \cite{biermann}, which works for
the generation of seeds to be amplified by the dynamo in
galaxies,
but can hardly account for the fields in clusters
\cite{clarke,bor}. Other astrophysical mechanisms (involving for instance
starbursts, jet-lobe radio sources, or accretion disks of black holes
in AGNs
\cite{colgate})
seem to need pre-existing magnetic fields. The current prevalent view is
that
the magnetic fields observed in galaxies and clusters have a
primordial origin (see \cite{grasso} for a review). The
processes that may account for this origin can be divided in two types:
causal (those in which the seeds are produced at a given time inside the horizon,
like QCD and EW phase transitions \cite{grasso}), and inflationary (where
correlations
are produced outside the horizon \cite{turner}). In the latter,
vacuum fluctuations of the electromagnetic field are
``stretched'' by the evolution of the background geometry
to super-horizon scales,
and they could appear today as large-scale magnetic fields.
However, since Maxwell's equations are conformally invariant
in the FRW background,
the amplification of the vacuum fluctuations (which amounts to particle production)
via inflation can work only if conformal invariance
is broken at some stage of the evolution of the universe. There are several ways in
which this invariance can be broken
\cite{dolgov2}: nom-minimal coupling between gravitation and
the electromagnetic field \cite{ns, turner, lamb}, quantum
anomaly of the trace of the stress-energy
tensor of electrodynamics \cite{dolgov1}, coupling of the EM field to a charged scalar
field \cite{turner, kandus}, exponential coupling between a scalar (whose potential drives
inflation) and the EM field \cite{ratra}, and a non-zero mass for the photon
\cite{proko}. Yet another possibility is
dilaton electrodynamics \cite{lemoine}, in which there is a scalar field
(the dilaton)
which couples exponentially to an Abelian gauge field. In this model
the inflationary expansion
is driven and not by the potential but by the kinetic term of the scalar field.
The exponential coupling is naturally implemented in the low-energy limit in string
theory \cite{copeland}, and
in Weyl integrable spacetime (WIST) \cite{wist}, and
can be viewed also as
a time dependence of the coupling constant, an idea
considered first by Dirac \cite{dirac}.
This avenue has been pursued
by Giovannini in a series of articles
\cite{gio1}.
More recently, a model
with both a dilaton and an inflaton was analyzed in \cite{bamba}, and
later generalized to a noncommutative spacetime \cite{bamba2}.
In all these articles, different aspects of photon production have been analyzed.
We will focus in this contribution on two definite items.
First ,
we shall address the creation of photons in a non-singular
universe described by an exact solution of the field equations.
More specifically,
in this solution
the passage through the bounce is described by the equations of the model
without resorting to unknown (Planck-scale) physics.
The results of this calculation are to be compared
with those of the second item, coming from a more complete model.
We shall take
into account in this second part the influence
of the
creation of matter in the squeezing of the vacuum state, a problem that
seems to have received
little attention in the literature.
It will be shown that as a result of
matter creation, there is an increment in the number of photons
originating in the stretching of the vacuum fluctuations. Thus this effect
may be of importance in the generation of primordial magnetic fields.
The analysis will be done in the
framework of
a phenomenological model, in which the creation of ultrarelativistic matter
is restricted to an interval centered in the bounce.
Two interesting features of the solution representing
this model are that it goes automatically into a radiation regime after a short time,
and it displays a constant value for the dilaton after entering the radiation era.
We shall start in Sect.1 by deducing a Hamiltonian for the EM field
in a curved background coupled to the scalar field. The quantization of this
Hamiltonian system will be presented in Sect.2, following the method
of the squeezed states. In Sect.3 we shall study the production of
photons in the absence of matter (\ie with only the scalar field
present). Variation of the photon number in a model that
takes into account matter creation will be discussed in Sect.4.
We close with a discussion in Sect.5.

\section{Field equations}

A time-dependent gauge coupling is a generic feature of 4-dimensional theories
obtained by compactification of some theory formulated in a spacetime
with more than 4 dimensions, such as Kaluza-Klein theory
\cite{love} and
string theory \cite{damour}. Time-dependent gauge couplings are also present in
WIST \cite{wist}. In all these cases, the action can be conveniently
written in the form
\begin{equation}
S=\half \int\,d^{4}x \sqrt{-g}\;
 f(\omega)  F_{\alpha\beta}\; F^{\alpha\beta},
\end{equation}
where  $\omega$ is the the dilaton (in the case of string theory)
or the scalar field associated to Weyl geometry (in the case of
WIST) and $F_{\mu\nu}$ is an Abelian field. The function
$f(\omega)$ will be left unspecified for the time being, but later
we will set $f(\omega) = {\rm e}^{-2\omega}$, which corresponds to
the case of string theory and WIST. The equations of
motion (EOM) that follow from this action are: \beq (f(\omega)
F^{\mu\nu})_{||\nu} = 0 ,
\label{eqB}
\eeq
\beq
\Fstar^{\alpha\beta}_{\;\;||\beta} = 0 , \label{eqE} \eeq where
the derivative wrt the background geometry is denoted by ''$||$'',
and the dual of the electromagnetic tensor $\Fstar_{\mu\nu}$ is
defined as
$$
\Fstar_{\alpha\beta} \equiv
\frac{1}{2}\eta_{\alpha\beta}\mbox{}^{\mu\nu} \, F_{\mu\nu}.
$$
Advantage will be taken in the following of the
formal equivalence of
Eqns.(\ref{eqB}) and (\ref{eqE}) and
the equations for the electromagnetic field in a material medium, an
equivalence that is fulfilled if we define
\beq
P^{\mu\nu} \doteq f(\omega)  F^{\mu\nu}.
\label{pmunu}
\eeq
A reference frame must be chosen in order to define the electric and magnetic
fields from $P_{\mu\nu}$ and $F_{\mu\nu}$. Since
the cosmological model
to be used in this article is described by the
Friedmann-Robertson-Walker (FRW) metric,
we can use the four velocity vector field $V^\mu = \delta
^\mu _{\;0}$, which is orthogonal to the three-dimensional surfaces of
homogeneity of the FRW geometry.
The electric and magnetic parts of the
electromagnetic tensor are defined as:
\beq
D^\alpha \doteq P^{\alpha\beta} \, V_\beta,
\;\;\;\;\;\;\;\;
 H^\alpha
\doteq \Pstar^{\alpha\beta} \, V_\beta,
\eeq
 \beq E^\alpha
\doteq F^{\alpha\beta} \, V_\beta,
\;\;\;\;\;\;\;\; B^\alpha
\doteq \Fstar^{\alpha\beta} \, V_\beta.
\eeq
In the comoving
frame, the EOM along with Eqn.(\ref{pmunu})
become
\beq
D^\alpha_{\;||\alpha}= 0, \label{eq.E1}
\eeq
\beq
B^\alpha_{\;||\alpha}= 0, \label{eq.B1}
\eeq
\beq
(h^{\;\alpha}_\lambda \, D_\alpha)^{\dot{}} + \frac{2}{3}
\, \theta \, D_\lambda +
\eta_{\lambda}\mbox{}^{\beta\rho\sigma} \, V_\rho \,
H_{\sigma ||\beta} =0 , \label{eq.E2}
\eeq
\beq
h^{\;\alpha}_\lambda \, \dot{B_\alpha} + \frac{2}{3} \, \theta  \,
B_\lambda + \eta_{\lambda}\mbox{}^{\beta\rho\sigma} \, V_\rho   \,
E_{\sigma||\beta} = 0 , \label{eq.B2}
\eeq
with $\theta = 3\dot a/a$, $a$ the scale factor of the FRW
metric, a dot represents derivative
wrt cosmological time, and $h^\alpha_\lambda$ is the metric of the 3-space
orthogonal to $v_\mu$.
As shown in the pioneering work by Lifshitz \cite{Lifshitz},
it is useful to expand the solutions of linear differential
equations defined on a curved manifold in the corresponding
spherical harmonics basis. In order to do so we will first split
the equations in space and time parts. The vectors describing the
electromagnetic field are already space vectors, since $
E_{\alpha}V^{\alpha}=D_{\alpha}V^{\alpha}=B_{\alpha}V^{\alpha}=H_{\alpha}V^{\alpha}=0$.
From now on we shall work in conformal time, with the FRW metric given by
$$
ds^2=a(\eta)^2\left[d\eta^2-\gamma_{ij}(\vec
x)dx{^i}dx{^j}\right],
$$
where $\gamma_{ij}$ is the metric of
the hypersurface $\eta=$constant, and $i=1,2,3$. In terms of 3-vectors,
Eqns.(\ref{eq.E1})-(\ref{eq.B2})
become
\beq
 \nabla_iD^i=0,
\label{eom1}
 \eeq
 \beq \nabla_iB^i=0,
\label{eom2}
 \eeq
\beq (D^i)'-\frac{a'}{3 a}D^i+\eta^{jln}\nabla_lH_n  =0,
\label{eom3}
\eeq
 \beq (B^i)'-\frac{a'}{3
a}B^i - \eta^{jln}\nabla_lE_n=0,
\label{eom4}
\eeq
where the prime denotes derivative wrt conformal time.

Since we are dealing with vector quantities we
will use a vector basis $P^i$ defined by the following relations
\cite{Klippert}
$$
P^{i}=P^i(x^i) ,
$$
$$
\gamma^{ij} \, \nabla_{i} \, \nabla_{j}\, P^l=-m^2 \, P^l,
$$
$$\gamma^{ij} \, \nabla_iP_j=0
$$
The eigenvalue $m$ denotes the wave number of the corresponding
vector eigenfunction of the Laplacian operator on the spatially
homogeneous hypersurfaces. The spectrum of eigenvalues depends on
the 3-curvature $\epsilon$, and is given by
$$
m^2 \, = \, s^2 \, + \, 2, \, 0<s<\infty,   \,
\epsilon=-1,
$$
$$
m \, = \, s, \, 0<s<\infty, \, \epsilon=0,
$$
$$
m^2\, = \, s^2 \, - \, 2, \, s=2,3,..,  \, \epsilon=1 .
$$\label{s3}
The pseudovector basis
$\Pstar^{i}$, convenient to work with the magnetic field, is defined by:
 \beq
 \Pstar^{i} = \eta^{ijl}  \, \nabla_lP_{j}.
  \eeq
In terms of the basis $P^i$
and of the associated basis $\Pstar^i$, the electric and magnetic
fields can be expanded as follows:
\beq
E^i(\eta,\vec x)= \sum_{l,m,\sigma} E^{(\sigma)}_{ml}(\eta) \, P^{(\sigma)
\, i }_{ml}(\vec x),  \label{elet}
 \eeq
 \beq
B^i(\eta,\vec x)=\sum_{l,\,m,\sigma} B^{(\sigma)}_{ml}(\eta) \,
\Pstar^{(\sigma)\, i}_{ml}(\vec x) . \label{magn}
 \eeq
These
expressions are valid in the case $\epsilon=1$. For
$\epsilon=0$ and $-1$, the sum must be replaced by an integration ($
(2\pi)^{-3}\int\,d^{3}x $). The index $l=1,2$ describes the two
transverse degrees of freedom of the electric and magnetic fields.
Since we are using an expansion in standing waves \footnote{Notice that
the expansion in standing waves is completely equivalent to that
of travelling waves (see for instance \cite{gri}).}, the index
$\sigma$ takes the values ''$+$'' or ''$-$''. Consequently, the fields
in
the case $\epsilon=0$
can be written as
 \beq E^i(\eta,\vec x)= \sum_{l,m}\left(
E^{(+)}_{ml}(\eta) \cos(\vec m\centerdot \vec x ) + E^{(-)}_{ml}(\eta)
\sin(\vec m\centerdot \vec x ) \right) e^i_l, \label{elet2}
\eeq
(where the $e^i_{\;l}$ are
modulus-one polarization vectors),
and an analogous expression for $B^i$. In the more general
case of $\epsilon=1, -1$ the (considerably more involved)
analytic expression for the vector
base can be found in \cite{kal}. The result of
substituting in Eqns.(\ref{eom1})-(\ref{eom4})
the expansions given in Eqns.(\ref{elet})
and (\ref{magn})
is
 \beq
 B^{(\sigma)'}_{ml}
+\frac{ a'}{a} \,  \, B^{(\sigma)}_{ml}-
f^{-1}(\omega)\,D^{(\sigma)}_{ml}=0, \label{ct1}
 \eeq
 \beq
 D^{(\sigma)'}_{ml} + \frac{ a'}{a} \, \,
D^{(\sigma)}_{ml} + (m^2+2\epsilon) \,
f(\omega)\,B^{(\sigma)}_{ml} = 0. \label{ct2}
 \eeq
The system described by
Eqns.(\ref{ct1}) and (\ref{ct2}) is not Hamiltonian when written
in the variables $D$ and $B$, but a Hamiltonian can be introduced
in terms of a new set of variables defined by
\beq
p^{(\sigma)}_{ml}(\eta) \equiv  a(\eta) D^{(\sigma)}_{ml}(\eta),
\eeq \beq q^{(\sigma)}_{ml}(\eta) \equiv a(\eta)
B^{(\sigma)}_{ml}(\eta).
 \eeq
In terms of these, Eqns.(\ref{ct1}) and (\ref{ct2}) can be written as
\beq
p^{(\sigma)'}_{ml} + (m^2+2\epsilon) f(\omega)\;
q^{(\sigma)}_{ml}=0,
\eeq
\beq
q^{(\sigma)'}_{ml}-
f(\omega)^{-1}\; p^{(\sigma)}_{ml}=0.
\eeq
Since the new variables constitute a pair of canonical variables, the
dynamical system given by Eqns.(\ref{ct1}) and  (\ref{ct2}) can
now be described by the Hamiltonian
\begin{eqnarray}
{\cal H}^{(\sigma)}_{ml}
(p,q)=\half\; b_2\; p^{(\sigma)2}_{ml} + \half\; b_1\;
q^{(\sigma)2}_{ml},
 \label{hamil}
\end{eqnarray}
where
\beq
b_1(\eta)= f(\omega)^{-1},\;\;\;\;\;
b_2(\eta)= (m^2+2\epsilon) f(\omega).
\label{b1b2}
\eeq
The Hamiltonian for the mode characterized by $(\sigma,m,l)$
given in Eqn.(\ref{hamil}) is identical to that of
a single harmonic oscillator problem with
time-dependent coefficients. We shall review in the next section
how to quantize this system.

\section{Quantization}

The expression for ${\cal H}$ obtained in the previous section
appears in many different branches of physics,
ranging from quantum optics
to gravitational waves \cite{schumi}.
In order to quantize it,
we shall follow a standard procedure of
quantum optics. First we define creation and annihilation
operators by the expressions
\beq
\hat{p}^{(\sigma)}_{ml}=-i\sqrt\frac{\gamma}{2}
\, \left(\hat{a}^{(\sigma)}_{ml}-\hat{a}^{(\sigma)\dagger}_{ml}\right), \eeq
\beq
\hat{q}^{(\sigma)}_{ml}=\frac{
\hat{a}^{(\sigma)}_{ml}+\hat{a}^{(\sigma)\dagger}_{ml}}{\sqrt{2\gamma}},
 \eeq
 where $\gamma$ is a constant to be determined later.
In terms of $\hat a$ and $\hat a^\dagger$,
the Hamiltonian operator can be written as
\beq
\hat{{\cal H}}^{(\sigma)}_{ml} = \fourth \;k(\eta) \;(2\hat{N}^{(\sigma)}_{ml} + 1) +
\fourth\; h (\eta)\; \hat{a}^{(\sigma)2}_{ml} +
\fourth \;h(\eta)\; \hat{a}^{(\sigma)\dagger 2}_{ml},
\label{ham2}
\eeq
where
$$
k(\eta) =\gamma b_1(\eta) + \frac{b_2(\eta)}{\gamma},
\;\;\;\;\;\;\;\;\;\;
h (\eta)  = \frac{b_2(\eta)}{\gamma}-\gamma b_1(\eta),
$$
and $
N^{(\sigma)}_{ml} =\hat{a}^{(\sigma)\dagger}_{ml} \, \hat{a}^{(\sigma)}_{ml}
$ is the number operator. The first term in Eqn.(\ref{ham2})
looks like the
free Hamiltonian for the mode $(\sigma,m,l)$ with a time-dependent
function instead of the frequency. This part conserves the number of
photons in each mode, hence the total number of photons and the
total energy. The second and third terms represent instead a
time-dependent interaction which does not conserve the photon number.

The operators $\hat{a}_{ml}^\dagger$ and $\hat{a}_{ml}$ obey the
commutation relations \beq
\left[\hat{a}^{(\sigma )}_{ml},\hat{a}^{(\sigma)'\dagger }_{m'l'}\right]= \delta_{ll'} \,
\delta_{mm'}\, \delta_{\sigma \sigma'}, \mbox{   for
$\epsilon=1$},
 \eeq
  \beq \left[\hat{a}^{(\sigma )}_{ml},\hat{a}^{(\sigma)'\dagger }_{m'l'}
\right]= \delta_{ll'} \,\, \delta_{\sigma\sigma'} \delta(m - m'),
\mbox{ for $ \epsilon= 0, -1$},
 \eeq
where we have settled $\gamma = m$ in order to have vacuum
as initial state.
Since we shall work with a single mode (and in
order to avoid clumsy expressions) the indices $m,l$ and $\sigma$
will be omitted from now onwards.

We will proceed to solve Schr\"odinger's equation for the
Hamiltonian given in Eqn.(\ref{ham2}) by writing the time
evolution operator as a product of the rotation and the
single-mode squeezed operators, along with a phase factor:
\beq
U(\eta,\eta_0)= e^{i\phi} \, S(\eta) \, R(\eta),    \label{evo}
\eeq
where $S(\eta)$ and $R(\eta)$ are defined as
\beq
S(\eta)=
e^{A(\eta) \, \hat{a}^2 - A^*(\eta) \, \hat{a}^{\dagger 2}},
\label{sque}
\eeq
\beq R(\eta)= e^{i\Gamma(\eta) \hat{a}^{\dagger}
\hat{a} }. \label{rot} \eeq The function $\Gamma(\eta)$ is called
the rotation angle, and $A(\eta)$ is usually defined as
\beq
A(\eta) = \frac{1}{2}\; r(\eta)\; e^{-2i\varphi(\eta)},
\label{param} \eeq where $r$ is the squeeze factor (defined in the
range $0\leq r < \infty$) and $\varphi$ the squeeze angle (defined
in the range $-\pi/2\leq\varphi<\pi/2$). The parameter $r$
determines the strength of the squeezing while $\varphi$ gives the
distribution of the squeezing between conjugate variables.
Schr\"odinger's equation with the Hamiltonian given in
Eqn.(\ref{ham2}) reduces, after a direct but somewhat long
calculation, to the following system of first order coupled
differential equations valid for each mode of the field:
 \beq
 ir'+(\varphi'
+\Gamma') \, \sinh 2r = e^{2i\varphi} \; \frac
{h}{ 2}, \label{e1}
\eeq
\beq
 -ir'+(\varphi' +
\Gamma') \, \sinh 2r = e^{-2i\varphi}\; \frac
{h}{ 2}, \label{e2}
\eeq
\beq (\varphi'  +
\Gamma') \, \cosh 2r - \varphi' = k,
\label{e3}
 \eeq
 \beq
\phi' =-\frac{\Gamma'}{2}. \label{e4}
 \eeq
Similar equations
were obtained by Albrecht \et \cite{Albrecht} in the case of the
scalar degrees of freedom of metric perturbations in cosmology.
Grishchuk \cite{gri2} derived analogous equations for the case of
gravitational waves, and Matacz \cite{Matacz}, for a scalar field.
To solve the system, we shall follow the method presented
in \cite{Matacz}. With the change of variables
\beq
\alpha(\eta) =
e^{-i\Gamma(\eta)}\cosh r(\eta), \label{alpha}
\eeq
\beq
\beta(\eta) = - e^{-2i(\varphi(\eta)+\Gamma/2)}\sinh r(\eta), \label{beta}
\eeq
Eqns.(\ref{e1})-(\ref{e4}) can be written as
\beq
2\alpha' =
-ih\beta-ik\alpha, \label{s1} \eeq
\beq
2\beta' = ik\beta + i h \alpha.
\label{s2}
\eeq
Introducing the function $\mu(\eta)$ defined by
\beq \mu =
\frac{\beta^*-\alpha^*}{\beta^*+\alpha^*}, \label{mudef}
\eeq
Eqns.(\ref{s1}) and (\ref{s2}) are equivalent to
\beq
2\mu' - i (k-h)\mu^2 +
i(k+h) =0. \label{mueq}
 \eeq
Finally, setting
\beq
 \mu  = \frac
{i}{\gamma b_1} \frac{g'}{g}, \label{mug}
 \eeq
we obtain the equation
$$
 g'' -
\frac{b_1'}{b_1} g' + b_1 b_2
g =0.
$$
The problem has been reduced then to solving a single second-order
differential equation for the function $g(\eta)$, which in terms of
$f(\omega) $ reads (see Eqn(\ref{b1b2}))
\beq
 g'' +
\frac 1 f \frac{df}{d\omega}\; \omega' g' + (m^2+2\epsilon)
g =0.
\label{gf}
\eeq
We can see from this equation that there is no photon creation
if the scalar field is constant, a fact that justifies the identification of
$\omega$ with the laser ``pump''used in experiments devised to observe squeezed
states \cite{schumi}.
With the function $g$ we
can get the squeeze parameter $r$
through Eqn.(\ref{mug}) as follows.
From Eqns (\ref{alpha}), (\ref{beta}), and (\ref{mudef}) we
get
\beq
\mu(\eta) = \frac{1+e^{2i\varphi(\eta)}\tanh
r(\eta)}{-1+e^{2i\varphi(\eta)}\tanh r(\eta)},
\label{mueta}
\eeq
which can be inverted to get $r$, given by
\beq \tanh ^2 r(\eta) =
\frac{1+\mu(\eta)+\mu^*(\eta)+|\mu(\eta)|^2}
{1-\mu(\eta)-\mu^*(\eta)+|\mu(\eta)|^2}. \label{r} \eeq
All
the functions appearing in the evolution operator can be written
in terms of
$\mu$ (we refer the reader to \cite{Matacz} for details).
A quantity of interest in the following is the mean number of photons
per mode as a function of the conformal time, given by \cite{giova152}
\beq
\langle N(\eta) \rangle = \sinh^2 r(\eta).
\eeq
In the next sections we shall calculate $\langle N \rangle$ for two different models.

\section{A simplified model}

\begin{figure}[h]
\begin{center}
\includegraphics[angle=-90,width=0.5\textwidth]{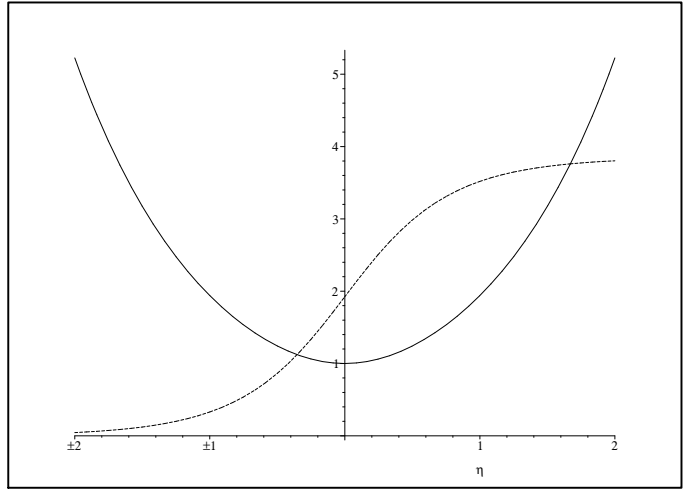}
\caption{Plot of the scale factor (full line) and $\omega$ as a function of the
conformal time for $\lambda=|b|=1$.}
\label{avacuo}
\end{center}
\end{figure}

\begin{figure}
\begin{center}
\includegraphics[angle=-90,width=0.5\textwidth]{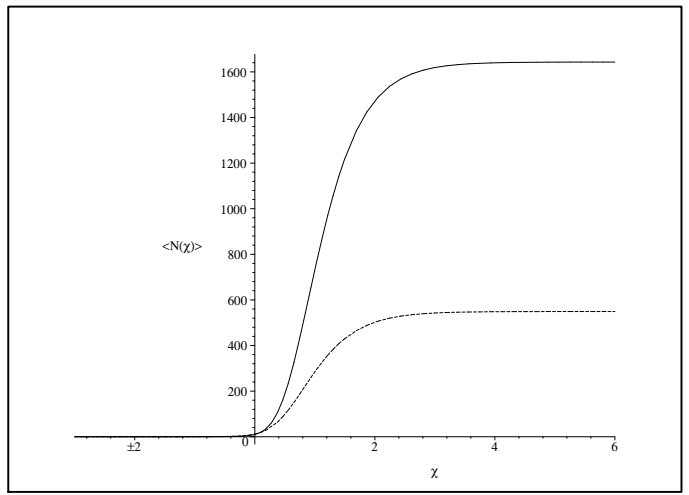}
\caption{Plot of the mean number of photons as a function of the conformal time
for $m=20$ for the case without matter
(dashed line) and
for the case with matter creation (full line, see next section), for
$\lambda =1$.}
\label{nvacuomateria}
\end{center}
\end{figure}

Before going into the details of the model
which takes into account the effect of
matter creation on the metric, we shall study the production
of photons in
a simpler case, namely that in which the sole matter content is
given by the scalar field $\omega$.
As discussed in \cite{Novello}, there are nonsingular solutions in
the theory of WIST that describe a FRW geometry plus a Weyl scalar
field. Nonsingular solutions are also present in string theory
\cite{massimo}.
Let us briefly review these solutions.
The EOM for gravitation plus scalar field written in conformal time are
\cite{Novello}
$$
a'^2 + \epsilon a^2 + \frac{\lambda^2}{6} (\omega'a)^2 =0,
$$
\beq
\omega' = \sigma a^{-2},
\label{energycons}
\eeq
where $\sigma =$ constant, and $\lambda^2$ is the coupling constant
of the scalar field to gravity.
Notice that the equation of state of the scalar field
is given by $\rho_\omega = p_\omega$, where
$$
\rho_\omega = -\frac{\lambda^2}
{2}\left(\frac{\omega'}{a}\right)^2.
$$
From these equations we get
\beq
a'^2 = -\epsilon  a^2 - \frac{b^2}{a^2},
\label{einstein1}
\eeq
where we have defined $b^2=\lambda^2\sigma^2/6$. Eqn.(\ref{einstein1})
shows that only solutions with $\epsilon =-1$ are possible. Hence, from now on
we shall restrict to the negative curvature case, for which
Eqn.(\ref{einstein1}) can be easily integrated.
Although this choice is at odds with observation, we emphasize
that we are using this exact
solution only as a toy model to study the effect of matter creation on
the production of photons.
The result of the integration for the scale factor is \footnote{The expression
for the scale factor in terms of cosmological time was given in \cite{Novello},
although not in a closed form}.
\beq
a(\eta) = \sqrt{|b|} \sqrt{\cosh (2\eta +\delta)}.
\eeq
From Eqn.(\ref{energycons}),
\beq
\omega(\eta) = \pm \frac{\sqrt{6}}{2\lambda} \arctan\left( e^{ 2\eta
+\delta}\right) + \frac\pi 4 \frac{sqrt 6}{\lambda},
\label{eqomega}
\eeq
where $\delta$ is an integration
constant.
The plots for these functions are given in Fig.\ref{avacuo}.
The scale factor displays a bounce, produced by the violation of the
strong energy condition by the scalar field \cite{matt}.
The number of photons created by the expansion in this model must be calculated numerically,
since Eqn.(\ref{gf}) has no analytical solution for the scalar field given in
Eqn.(\ref{eqomega}) with the coupling $f(\omega)=e^{-2\omega}$.
The results are given by the
dashed curve of
Fig.(\ref{nvacuomateria}), for a wavenumber typical of the size
of the intergalactic scale (1 Mpc).

\section{A model with matter creation}

\begin{figure}
\begin{center}
\includegraphics[angle=-90,width=0.5\textwidth]{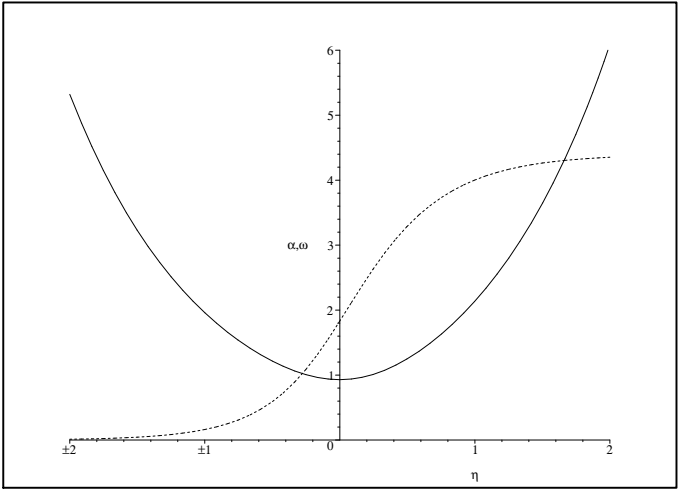}
\caption{Plot of $a$ and $\omega$ for $k_0=1/7$ and $a_0=0.93$,
values chosen by imposing that the solution in Eqn.(\ref{anosso})
enters the radiation era for $t\approx 10^{-8}$ seg.}
\label{aomegamatter}
\end{center}
\end{figure}

\begin{figure}
\begin{center}
\includegraphics[angle=-90,width=0.5\textwidth]{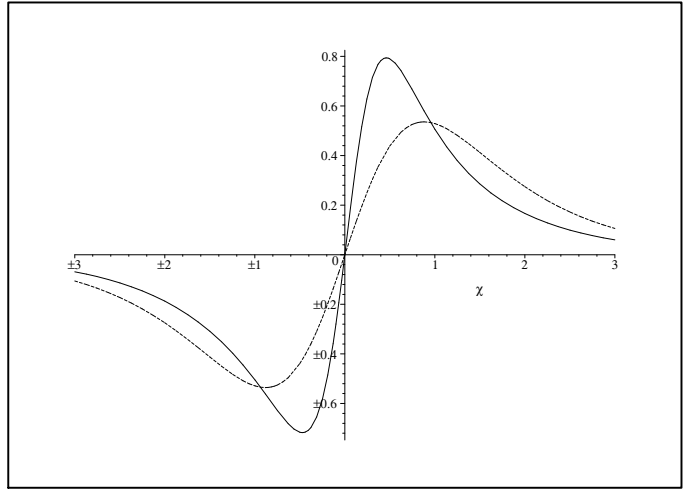}
\caption{Plot of $H$ for the vacuum case (full line)
and for the solution given by Eqn.(\ref{anosso})
as a function of the conformal time for $\lambda =1$ in $c=1$
units.}
\label{hubble}
\end{center}
\end{figure}

We shall consider in this section the production of photons
in a background of the system composed of the scalar field
plus matter and geometry, using a non-singular new solution that
incorporates the effect of the creation of matter on the geometry.
Friedmann's equation in conformal time for this case is given by
\beq
a'^2 - {a^2} = -\frac{\lambda^2}{6}(\omega'a)^2+ \frac{a^4}{3} \rho_m ,
\label{einstein2}
\eeq
while the second Einstein equation is
\beq
-3\left(2\frac{a''}{a^3}-\frac{a'^2}{a^4}-\frac{1}{a^2}\right) =
\rho_m + 3\rho_\omega.
\label{2ee}
\eeq
The conservation of the stress-energy tensor gives
\beq
\frac{d}{d\eta}(a^3(\rho_m + \rho_\omega))
+ (p_m + p_\omega)\frac{da^3}{d\eta}    = 0.
\eeq
In the case of ultrarelativistic matter, this equation takes the form
\beq
\left( a^4\rho_m\right)'+ \frac{1}{a^2}\left(a^6\rho_\omega\right)'=0.
\label{ec}
\eeq
We would like to have a solution that describes creation of relativistic matter
only around the bounce, and enters a radiation phase
with a constant scalar field
in a short time.
Clearly, an asymmetry is to be expected
both in the scale factor and in $\omega$,
since the evolution of this universe starts from the vacuum and enters
a radiation-dominated epoch. An expression for
$a$ that fulfills these requirements is
\beq
a(\eta) = \beta \sqrt{\cosh(2\eta)+k_0\sinh(2\eta)-2k_0
(\tanh(\eta)+1)},
\label{anosso}
\eeq
with $\beta = a_0\sqrt{\frac{1}{1-2k_0}}$ and $0<k_0\leq 1/7$.
The expression for $\omega$ can be obtained using this scale factor in
Eqns.(\ref{einstein2}) and (\ref{2ee}). We shall not give the explicit expression, but
the plots for $a(\eta)$ and $\omega(\eta)$ (see Fig.\ref{aomegamatter}).
Notice that the plot shows the announced asymmetry.
The evolution of the Hubble parameter for the vacuum case
(studied in the previous section) and for the case with matter
creation is showed in Fig.\ref{hubble}.
Let us emphasize that since the scalar field tends rapidly to a constant value,
the production of matter (controlled by $\omega'$, see
Eqn.(\ref{ec})) stops soon,
and the model enters a radiation phase
\emph{without the need of a potential}. In this sense, this solution describes a
\emph{hot bounce}, as opposed to cold bouncing solutions, which do not enter
the radiation era unless they are heated up \cite{giovacqg}. Another nice feature of
this solution is that the dilaton goes automatically to a constant value for
$\eta\rightarrow\infty$, in such a way that the solution could be taken as the leading
order of a perturbative development (as is the case in string theory).
Again, no potential was needed in order to
display this feature.
The mean number of photons per mode can be calculated as in the previous cases, after
numerical integration of Eqn.(\ref{gf}).
The full line plot in Fig.\ref{nvacuomateria} shows the that
the production of photons is increased in the case of the
model with matter creation. The number of photons is directly related to
the fraction of electromagnetic energy stored per mode, relative to the background radiation
energy $\rho_\gamma$ through the expression
$$
r(\nu)\approx \frac{\nu^4}{\rho_\gamma}\langle N\rangle_\nu,
$$
where $\nu=\sqrt{m^2-2}/a$. Consequently, the
increment in the number of photons may be of importance in the problem of the generation of
magnetic field seeds.

\section{Discussion}

Using a covariant description for the electromagnetic field coupled to a scalar
in a curved background, we obtained the Hamiltonian for each mode of the field.
With the aid of the formalism of the squeezed states,
we have calculated the number of photons in a non-singular universe using first
an already known solution \cite{Novello} with no matter present,
and afterwards a new solution which takes into account the effect of the creation of
ultrarelativistic matter on the evolution of the scale factor and of the scalar field.
This new solution presents several interesting features, namely the
transition of an expanding phase to a radiation era, and the constancy of the
scalar field a short time after the bounce.
The results for the mean photon number per mode show that the production is
increased in the case of matter production. This increment may be relevant
for the creation of seeds of the magnetic field. We hope to discuss in detail
this issue in a forthcoming publication.

\section*{Acknowledgements}

SEPB acknowledges UERJ for financial support. JS is supported by CNPq.


\begin{thebibliography}{99}



\bibitem{parker} E. Parker, Ap. J. {\bf 162}, 665 (1970).
See also
Y. B. Zeldovich, A. A. Ruzmaikin, and D. D. Sokoloff,
\emph{Magnetic Fields in Astrophysics}, Mc Graw-Hill, New York (1980).

\bibitem{kul} See however
R. Kulsrud, S. C. Cowley, A. V. Gruzinov and R. N. Sudan, Phys. Rep.
{\bf 283}, 213(1997), and L. M. Widrow, Rev. Mod. Phys. {\bf 74}, 775 (2003).


\bibitem{biermann} L. Biermann, Z. Naturf. 5A, 65 (1950).


\bibitem{clarke} T.E. Clarke, P.P. Kronberg and H. B\"ohringer, Astrophys. J. {\bf 547},
L111
(2001).

\bibitem{bor} H. B\"ohringer, Rev. Mod. Astron. {\bf 8}, 295 (1995).


\bibitem{colgate} S. A. Colgate and H. Li, astro-ph/0001418.

\bibitem{grasso} D. Grasso and H. Rubinstein, Phys. Rept. {\bf 348}, 163 (2001).


\bibitem{turner} M. Turner and L. Widrow,
Phys. Rev. D {\bf 37}, 2743 (1988).


\bibitem{dolgov2} For a review on different mechanisms of seed generation, see
\emph{Generation of magnetic fields in cosmology}, A. Dolgov,
in Gurzadyan, V.G. (ed.) et al.: \emph{From integrable models to gauge theories},
143-154, \texttt{hep-ph/0110293}.

\bibitem{lamb} \emph{Gauge invariant wave equations in curved space-times and
primordial magnetic fields}, G. Lambiase and A. Prasanna,
\texttt{gr-qc/0407071}.


\bibitem{ns} M. Novello and J. Salim, \Prd {\bf 20}, 377 (1978).

\bibitem{dolgov1} A. Dolgov, \Prd {\bf 48}, 2499 (1993).

\bibitem{kandus} E. Calzetta, A. Kandus and F. Mazzitelli,
Phys. Rev. D {\bf 57}, 7139 (1998).


\bibitem{ratra} B. Ratra, Astrophys. J. {\bf 391}, L1 (1992).

\bibitem{proko} \emph{Nearly minimal magnetogenesis}, T. Prokopec and
Ewald Puchwein, \texttt{astro-ph/0403335}.

\bibitem{lemoine}
Lemoine D., Lemoine M., Phys. Rev. {\bf D 52}, (1995),
1955, Gasperini M., Giovannini M. and Veneziano G. Phys. Rev.
Lett. {\bf V 75}, (1995), 3796.



\bibitem{copeland} J. Lidsey, D. Wands, E. Copeland,
Phys. Rept. {\bf 337}, 343 (2000), \texttt{hep-th/9909061}.

\bibitem{wist} J. M. Salim and S. Sautu, Class. Quant. Gravity, {\bf 13} (1996) 353-360,
J. M. Salim, S. Sautu and Martins R., Class. Quant. Gravity,
{\bf 15} (1998) 1521.



\bibitem{dirac} P. Dirac, Nature {\bf 139}, 323 (1937).

\bibitem{gio1} See \emph{Magnetogenesis, variation of gauge couplings,
and inflation}, M. Giovannini,
Proceedings of the Chalonge School on Astrofundamental Physics,
N.G. Sanchez and Y.M. Pariiski, eds. Kluwer (2002),
\texttt{astro-ph/0212346}.

\bibitem{bamba} K. Bamba
and J. Yokoyama, \Prd {\bf 69}, 043507 (2004),
\texttt{astro-ph/0310824}.

\bibitem{bamba2} \emph{Large-scale magnetic fields from dilaton inflation in
noncommutative spacetime}, K. Bamba
and J. Yokoyama, \texttt{hep-ph/0409237}. See also \emph{Generation of Cosmological seed magnetic
fields from inflation with cutoff}, A. Ashoorioon and R. Mann, \texttt{gr-qc/0410053}.

\bibitem{love} See for instance D. Bailin and A. Love,
Rept. Prog. Phys. {\bf 50}, 1087 (1987).

\bibitem{damour} See for instance T. Damour and  A.
M. Polyakov, Nucl. Phys. {\bf B423}, 532 (1994), \texttt{hep-th/9401069}.

\bibitem{Lifshitz}
Lifshitz E. M. and Khalatnikov I. M., Adv. Phys. {\bf 12},
(1963), 185.

\bibitem{Klippert}
M. Novello, J. M. Salim, M.C. Motta da Silva, S. E. Joras and R.
Klippert, Phys. Rev. \ {\bf D 52}, (1995), 730.

\bibitem{kal} E. Kalnins and W. Miller, Jr., J. Math. Phys {\bf 32}, 698 (1991).

\bibitem{schumi} B. Schumaker, Phys. Rep. {\bf 135}, 317 (1986).

\bibitem{Albrecht} A. Albrecht, P. Ferreira, M.
Joyce, T. Prokopec, Phys. Rev. D {\bf 50}, 4807 (1994),
{\tt astro-ph/9303001}.


\bibitem{gri} L. Grischuk and Y. Sidorov, \Prd {\bf 42}, 3413 (1990).

\bibitem{gri2} L. Grishchuk, \Prd {\bf 45}, 4717 (1992).

\bibitem{Matacz}
Matacz A. L., Phys. Rev. {\bf D 49}, (1994), 788.

\bibitem{Novello}
M. Novello, L. Oliveira, J.M. Salim and E. ELbaz, Int. J. of
Mod. Phys. {\bf D 1}, 641 (1993).


\bibitem{massimo} See for instance \emph{Old ideas and new twists in string cosmology},
M. Giovaninni, \texttt{hep-th/0409251}.

\bibitem{giova152}M. Giovannini, \emph{Primordial Magnetic Fields},
\texttt{hep-ph/0208152}.







\bibitem{giovacqg} M. Giovannini, \Cqg {\bf 21}, 4209 (2004).


%
%
%
%
%
%
%
%
%
%
%
%
%
%
%
















%
%
%
%
%
%
%
%
%
%
%
%
%





\bibitem{matt} C. Molina-Paris and M. Visser \Plb {\bf 455}, 90 (1999).

\end{thebibliography}
\end{document}